\documentclass[superscriptaddress,aps,floatfix,prl,10pt]{revtex4-2}
\usepackage{hyperref}
\usepackage{blindtext}
\usepackage{amsmath}
\usepackage{graphicx}
\usepackage{mathrsfs}
\usepackage{braket}
\usepackage{color}
\usepackage{multirow}

\renewcommand{\Im}{\operatorname{Im}}

\begin{document}
	\title{Topological Single Photon Emission from Quantum Emitter Chains}
	
	\author{Yubin Wang$^\ddagger$}
	\affiliation{State Key Laboratory of Low-Dimensional Quantum Physics and Department of Physics, Tsinghua University, Beijing, 100084, P.R. China}
	\affiliation{Beijing Academy of Quantum Information Sciences, Beijing, 100193, P.R. China}
	
	\author{Huawen Xu$^\ddagger$}
	\affiliation{Division of Physics and Applied Physics, School of Physical and Mathematical Sciences, Nanyang Technological University, Singapore, 637371, Singapore}
	
	\author{Xinyi Deng}
	\affiliation{Beijing Academy of Quantum Information Sciences, Beijing, 100193, P.R. China}
	
	\author{Timothy Liew}
	\affiliation{Division of Physics and Applied Physics, School of Physical and Mathematical Sciences, Nanyang Technological University, Singapore, 637371, Singapore}
	
	\author{Sanjib Ghosh}
	\email{sanjibghosh@baqis.ac.cn}
	\affiliation{Beijing Academy of Quantum Information Sciences, Beijing, 100193, P.R. China}
	
	\author{Qihua Xiong}
	\email{qihua{\_}xiong@tsinghua.edu.cn; \\ $^\ddagger$ These authors contributed equally to this work}
	\affiliation{State Key Laboratory of Low-Dimensional Quantum Physics and Department of Physics, Tsinghua University, Beijing, 100084, P.R. China}
	\affiliation{Beijing Academy of Quantum Information Sciences, Beijing, 100193, P.R. China}
	\affiliation{Frontier Science Center for Quantum Information, Beijing, 100084, P.R. China}
	\affiliation{Collaborative Innovation Center of Quantum Matter, Beijing, P.R. China}
	
	\begin{abstract}
		We develop a scheme of generating highly indistinguishable single photons from an active quantum Su-Schrieffer-Heeger chain made from a collection of noisy quantum emitters. Surprisingly, the single photon emission spectrum of the active quantum chain is extremely narrow compared to that of a single emitter or topologically trivial chain. Moreover, this effect becomes dramatically strong close to the non-trivial-to-trivial phase transition point. Using this effect, we show that the single photon linewidth of a long topological quantum chain can become arbitrarily narrow, constituting an ideal source of indistinguishable single photons. Finally, taking specific examples of actual quantum emitters, we provide a microscopic and quantitative analysis of our model and analyze the most important parameters in view of the experimental realization.
	\end{abstract}
	
	\maketitle

	\textit{\textbf{Introduction:}} Single photon sources with high purity, indistinguishability, and efficiency are key building blocks of the rising quantum technologies~\cite{obrien_photonic_2009}. Among the different approaches, heralded photon generation is based on nonlinear processes, such as atomic cascade \cite{PhysRevD.9.853}, parametric down-conversion \cite{PhysRevLett.56.58}, and four-wave mixing \cite{Sharping:01}, where detection of one from a pair of photons heralds the presence of the other photon. However, such processes are stochastic and the efficiency is typically low. Alternatively, an important approach to single photon generation is based on the quantum emitters, which are effectively two-level quantum systems with the ability to emit one photon at a time. There are several promising systems for realizing ideal single photon emitters, such as single atoms \cite{PhysRevLett.39.691}, single ions \cite{PhysRevLett.58.203}, single molecules \cite{PhysRevLett.69.1516}, color centers \cite{Brouri:00}, and semiconductor quantum dots (QDs) \cite{lounis_photon_2000}. Intriguingly, microscopic defects \cite{srivastava_optically_2015, he_single_2015, koperski_single_2015, chakraborty_voltage-controlled_2015, tran_quantum_2015} and moir\'{e}-trapped excitons \cite{yu_moire_2017, baek_highly_2020} in two-dimensional materials and perovskite colloidal QDs \cite{park_room_2015, hu_superior_2015, utzat_coherent_2019, huo_optical_2020} are also demonstrated to emit single photons. Several of these systems promise efficient on-demand single photon emission with scalability and on-chip integration \cite{PhysRevLett.116.020401, somaschi_near-optimal_2016, senellart_high-performance_2017, PhysRevLett.122.113602, wang_towards_2019}.
	
	While various semiconductor QDs remain as viable systems as quantum emitters, a serious challenge comes from the large spectral linewidth. It severely diminishes the  single photon indistinguishability \cite{PhysRevLett.104.137401, PhysRevLett.116.033601}, which is crucial for quantum communications, quantum computing, and quantum metrology. Such a linewidth broadening is typically attributed to the random energy fluctuation caused by the spectral diffusion \cite{PhysRevLett.77.3873, PhysRevB.61.R5086, ham_composition-dependent_2018} and phonon scattering \cite{gammon_homogeneous_1996, besombes_acoustic_2001}. The spectral diffusion results from a time-fluctuating quantum confined Stark effect due to the interaction with charged defects in the electrostatic environment \cite{empedocles_s_a_quantum-confined_1997, sharma_stark_2019}, which strongly reduces the indistinguishability of the single photon emission. The spectral diffusion can be partially suppressed by applying sequences of optical pulses \cite{PhysRevLett.116.033603}, strong illumination \cite{liu_optical_2017}, and using the Purcell effect induced by a photonic cavity \cite{lyasota_limiting_2019} or a Fano cavity-photon interface \cite{huang_fano_2022}. However, all these methods require external controls, which can seriously constrain the scalability and have limited efficiency in suppressing spectral diffusion. The phonon scattering, resulting from exciton-acoustic phonon coupling, accounts for the homogeneous broadening of the single photon emission. This effect becomes stronger as temperature increases \cite{ota_temperature_1998, bayer_temperature_2002}, which impedes single photon emission at room temperature.
	
	An intriguing route to gain intrinsic robustness in photonic systems is to introduce non-trivial topology \cite{lu_topological_2014}. However, most of the existing topological systems operate in the classical or semi-classical regime \cite{hafezi_imaging_2013, rechtsman_photonic_2013,ota2020active,ozawa2019topological}, where the microscopic quantum nature of photons plays no significant role. For instance, topological insulators were used for robust lasing in the edge modes extending over many lattice sites along the boundaries of the systems\cite{harari_topological_2018, bandres_topological_2018, zeng_electrically_2020, zhao2018topological, shao2020high}. The robustness in the lasing is obtained from the global topology of the edge modes, which detours the local defects and deformations through the nearby alternative paths, thereby keeping the global topology unaffected. Such methods do not naturally fit into the zero-dimensional geometry of quantum emitters like QDs. Although strong optical nonlinearity can be utilized for single photon generation, it is typically local in nature, and thus extended edge states of the topological insulators would be incompatible \cite{verger_polariton_2006, ghosh_dynamical_2019}. Indeed, topological structures are typically used as passive platforms to transport quantum information \cite{barik_topological_2018, mittal_topological_2018}. However, active quantum systems like single photon emitters must also be robust to realize an infallible quantum platform ultimately. Here, we investigate the possibility of using the Su-Schrieffer-Heeger (SSH) topology in active quantum systems to generate robust single photons. Since the emission has to be localized at a single site and not from an entire edge state, it is counter-intuitive to expect topological robustness. Indeed, we find that the spectrum of such a system, in general, is no better than that of an isolated quantum emitter. Surprisingly, the scenario completely changes near the topological non-trivial-to-trivial phase transition point, where the single photon emission shows dramatic robustness to noise. Here, the single photon spectrum becomes arbitrarily narrow for a large system size. This provides highly indistinguishable single photons even in a strongly noisy environment. We provide the microscopic mechanism and explore physical systems for experimental realization.
	
	~\\
	
\begin{figure}[]
		\centering
		\includegraphics[width= 0.48\textwidth]{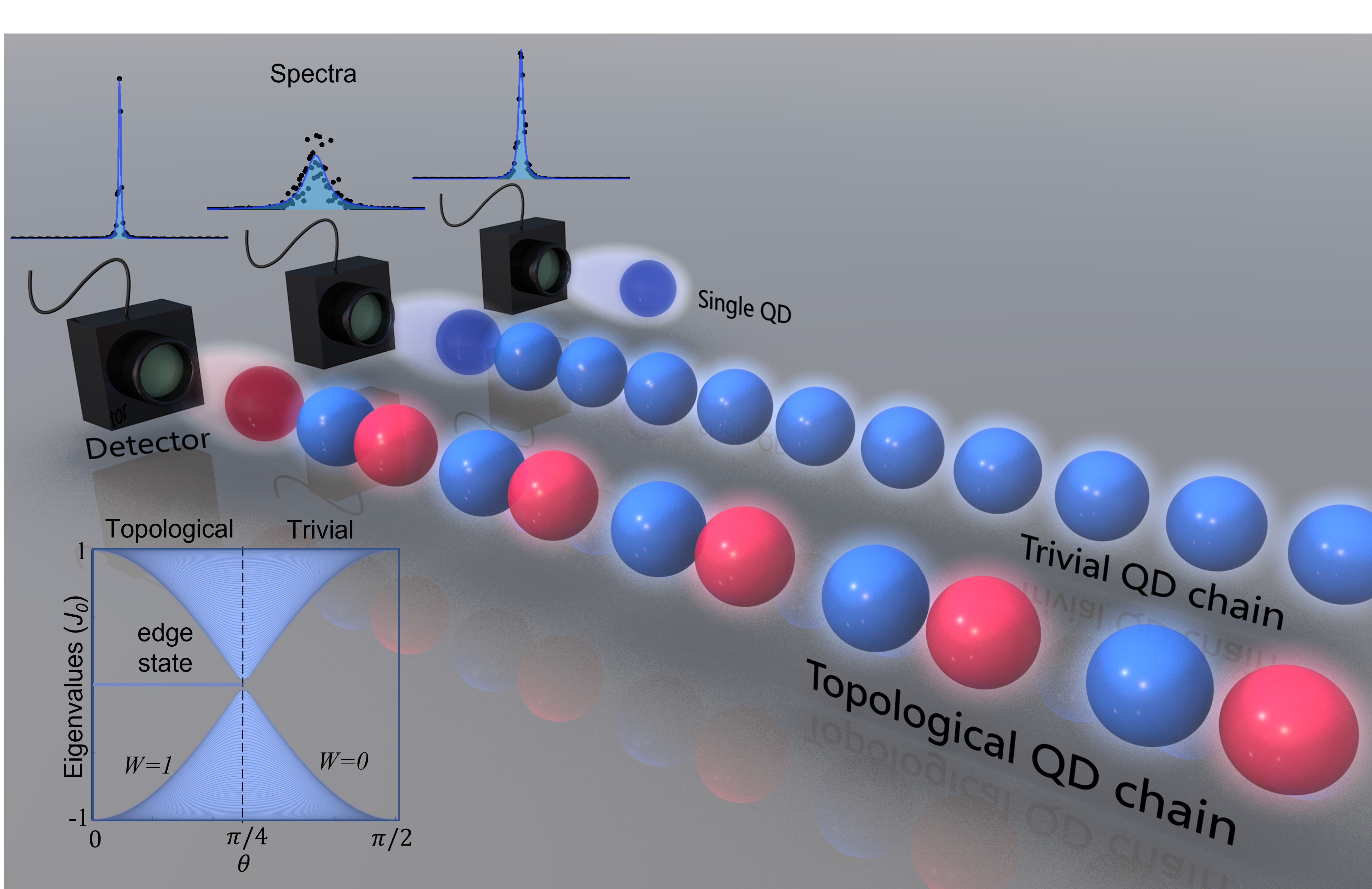}
		\caption{Three schemes of generating single photons: single QD, topologically trivial and non-trivial chains of QDs. We find that while a trivial chain broadens the spectrum (compared to a QD), a topologically non-trivial chain narrows the spectrum. Inset: Eigenvalues of the SSH model are plotted with the parameter $\theta=\arctan(\sqrt{J/J'})$. The topological non-trivial to trivial phase transition occurs at $\theta=\pi/4$.} 
		\label{Fig.1}
	\end{figure}

	\textit{\textbf{Models:}} A set of quantum emitters (two-level quantum systems) arranged in an SSH chain~\cite{Su1980,Heeger1988} constitutes an active quantum system, which can be described by a Hamiltonian $\mathcal{H}$:
	\begin{equation}
		\mathcal{H} = \sum_j\varepsilon_j(t)\sigma_j^+\sigma_j^-+\sum_jJ_j(\sigma_j^+\sigma_{j+1}^-+\sigma_{j+1}^+\sigma_j^-),
	\end{equation}
	where $\sigma_j^+$ ($\sigma_j^-$) is the creation (annihilation) operator associated with the QD emitter at the site $j$. Here, the QD is defined as a two-level quantum system with the ability to emit one photon at a time. The hopping amplitude $J_j=J$ for odd $j$ and $J_j=J^{\prime}$ for even $j$. The onsite energy $\varepsilon_j(t)$ represents microscopic energy fluctuations: $\varepsilon_j(t)=\epsilon\delta\varepsilon_j(t)$, where $\delta\varepsilon_j(t)$ is a Gaussian random variable and $\epsilon$ is the noise strength. The noise could originate from time-fluctuating environmental effects such as charge defects and phonon scattering. In the ideal condition, the SSH chain with the onsite energy $\varepsilon_j(t)=0$ (or a constant value) has a winding number:
	\begin{equation}
		W=\frac{1}{2\pi i}\int_{-\pi}^{\pi}dk\frac{d}{dk}log[h(k)],
		\label{EqWinding}
	\end{equation}
	where $h(k)=J+J^{\prime}e^{ik}$. We parameterize $J=J_0\sin^2\theta $ and $J'=J_0\cos^2\theta$, where $J_0$ is a constant and $\theta$ is an angular parameter ranging in $[0,\pi/2]$. For $\theta<\pi/4$, the chain has a non-trivial topology with a winding number $W=1$. Our idea is to obtain single photons from the topological edge states (see Fig.~\ref{Fig.1}), which are energetically located inside a band gap.
	
	Here we focus on the spontaneous emission spectrum $S(\omega)$ which is related to the time-dependent many-body correlation function:
	\begin{equation}
		C_j(t,t')=\langle {\sigma_j^+(t)\sigma_j^-(t^{\prime})} \rangle,
	\end{equation}
	where $\braket{\mathcal{O}}$ indicates the expectation value of an operator $\mathcal{O}$. We aim to find the emission properties of the system with the underlying non-trivial topology. For this, we consider that the system is put into an excited state at time $t=0$, and the emission from an edge site is detected by an ideal detector, which records photons for an infinitely long time. The corresponding single photon spectrum is given by \cite{PhysRevA.40.5516}
	\begin{equation}
		S(\omega)= \frac{\int_0^\infty dt\int_0^\infty dt' \,C_j(t,t')^\prime e^{-i\omega(t-t^\prime)} }{2\pi\int_0^\infty dt\, C_j(t,t)}.
	\end{equation}
	For instance, the spectrum of an ideal emitter with a long lifetime is given by $\delta(\hbar\omega-\varepsilon)$, i.e., the emitted photons are indistinguishable with a fixed energy $\hbar\omega=\varepsilon$. We note that the lifetime of a quantum emitter is of the order of nano-second which corresponds to a negligibly smaller linewidth compared to the actual linewidth, see Table S1 in Supplementary Information (SI).  
	
	~\\
	
\begin{figure}[h!]
		\centering
		\includegraphics[width=0.48\textwidth]{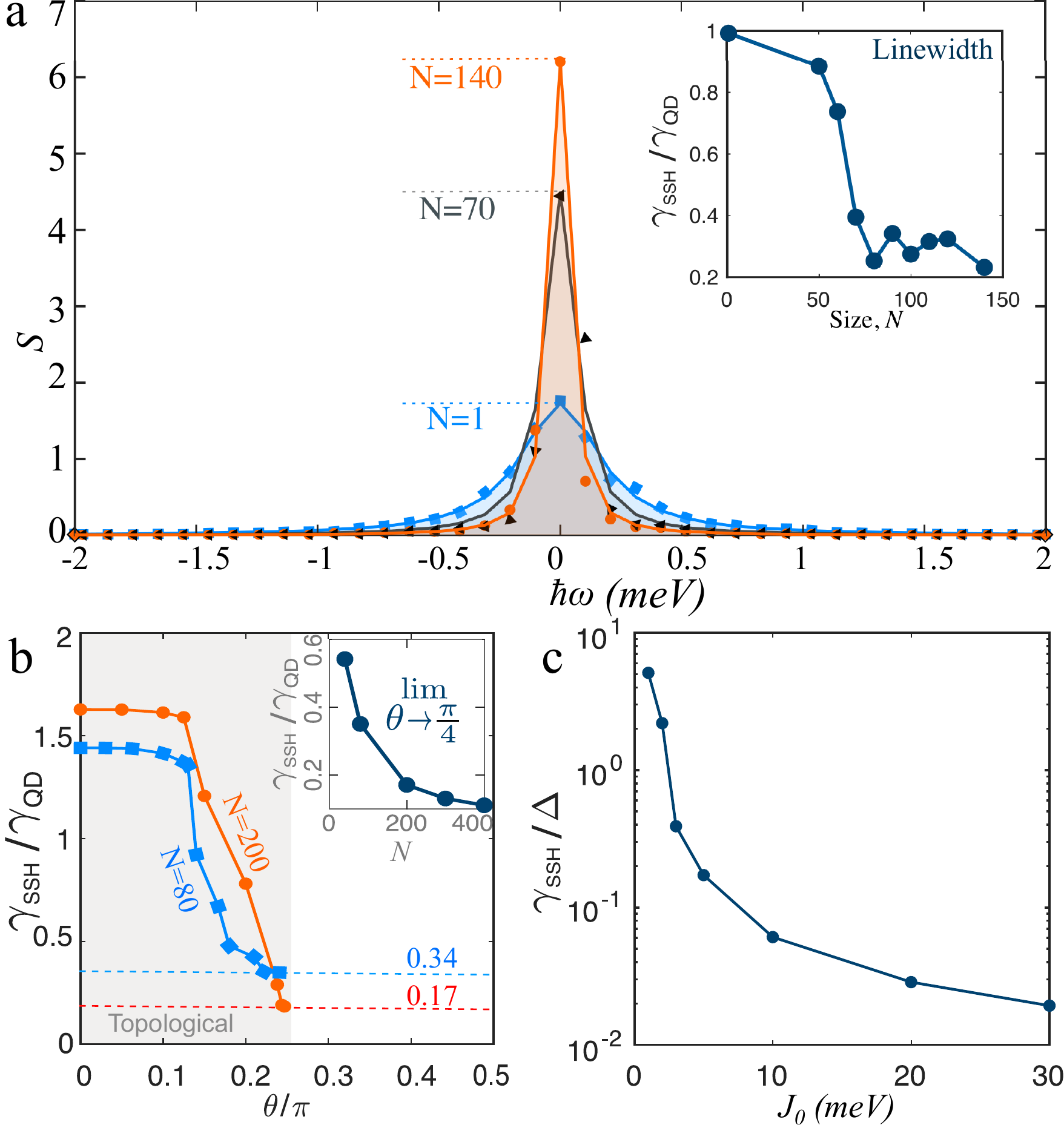}
		\caption{Topological linewidth narrowing. (a) The single-photon spectrum $S(\omega)$ for a topological SSH chain ($\theta=\pi/4.2$) of size $N$. The points and lines are the data and corresponding Lorentzian fits respectively. The inset shows the spectral linewidth $\gamma_\text{SSH}$ (full-width-at-half-maximum) as a function of $N$ at $\theta=\pi/4.2$. For an isolated QD ($N=1$), we numerically find that the linewidth $\gamma_\text{QD}\approx 0.4\, meV$. (b) The variation of $\gamma_\text{SSH}$ (normalised by $\gamma_\text{QD}$) with $\theta$ for $N=80$ (blue squares) and $200$ (red dots). The linewidth dramatically decreases near $\theta =\pi/4$. As indicated by the horizontal dotted lines, the value $\gamma_\text{SSH}(\theta\to \pi/4)$ becomes smaller for larger $N$. Inset: $\gamma_\text{SSH}(\theta\to \pi/4)$ becomes systematically smaller with the increasing $N$. (c) The ratio between $\gamma_\text{SSH}$ and the band gap $\Delta$ as a function of the hopping amplitude $J_0$. We consider $\epsilon=0.5\,meV$, $\tau=0.5\,ps$, $\theta=\pi/4.2$, and $J_0=30\,meV$ when they are not used as variables. For simulation, we use a topological edge mode as the initial state and $10$ noise realizations for average.}
		\label{TopoNarrowing}
	\end{figure}

	\textit{\textbf{Results:}} The spectrum of an isolated emitter can be obtained from the retarded Green's function: $S(\omega) = -\Im[G^R(\omega)]/\pi$. For an ideal emitter, the Green's function is given by $G^R(\omega) = lim_{\eta\to 0}(\omega\hbar+i\eta)^{-1}$ which implies a single energy emission at $\hbar\omega=0$. In the presence of noise, the system can be described by an average Green's function: $\overline{G}^R(\omega)=\lim_{\eta\rightarrow0}[\omega\hbar+i\eta-\Sigma]^{-1}$ where $\Sigma$ is the self-energy of the emitter~\cite{Mahan1990}. The corresponding spontaneous emission spectrum is given by
	\begin{equation}
		S(\omega)=\frac{1}{\pi}\frac{-\Im\Sigma}{(\hbar\omega)^2+(\Im\Sigma)^2},
	\end{equation}
	which is a Lorentzian distribution with the linewidth (full width at half maximum): $\gamma_\text{QD}\approx -2\Im\Sigma$. It is determined by the noise correlation function: $\overline{\delta\varepsilon_j(t)\delta\varepsilon_j(t^\prime)}=\epsilon^2e^{-(t-t^\prime)^2/(2\tau^2)}$ where $\tau$ is the correlation time, and the over-line represents the noise average. In the lowest order Dyson series~\cite{Mahan1990} $\Sigma \approx -(i\epsilon^2\tau\sqrt{\pi})/(\hbar\sqrt{2})$. We find that while the exact numerical value $\gamma_\text{QD} = 0.400\, meV$ (see Fig.~\ref{TopoNarrowing}), the analytical values are $0.470\, meV$, and $0.403\,meV$ with the lowest order approximation, and a self-consistent method respectively, see SI for details. This non-zero linewidth signifies the so-called spectral diffusion, e.g., an emitted photon will have a slightly different energy each time. This poses a serious challenge for application in quantum technologies \cite{doi:10.1080/09500340.2012.687500}.
	
	~\\
	
	\textit{Topological linewidth narrowing:} In Fig. \ref{TopoNarrowing}, we show the linewidth narrowing of single photon emission by coupling QDs in a topologically non-trivial SSH chain. The emission is considered at the edge of the SSH chain. We show the emission spectrum for different sizes of the SSH chain. Surprisingly, as the length of the chain $N$ increases, the linewidth $\gamma_\text{SSH}$ (full width at half maximum) drastically decreases, see Fig.~\ref{TopoNarrowing}a. This suggests that the underlying topological non-triviality greatly reduces the effective noise acting on the edge QD. However, this effect is weak for $J/J'\ll 1$ ($\theta\ll \pi/4$) even though it is within the regime of topological non-triviality. This can be understood from the fact that for $J\ll J'$, the coupling between the edge QD and the rest of the SSH chain is weak, and thus the edge QD is practically isolated. The effect of the global topology in the chain becomes significant near the phase transition point $\theta=\pi/4$ where $J/J'\sim 1$. In Fig.~\ref{TopoNarrowing}b, we show that the linewidth $\gamma_\text{SSH}$ becomes dramatically small near $\theta =\pi/4$. Importantly, the linewidth narrowing becomes stronger as the system size $N$ increases. In fact, it can be made arbitrarily stronger by choosing a large $N$ and $\theta \to\pi/4$ (see Fig.~\ref{TopoNarrowing}b). However, the topological robustness of a system is ensured only if the linewidth is much smaller than the band gap $\Delta$, i.e., $\gamma_\text{SSH}/ \Delta \to 0$. Since $\Delta$ is proportional to the hopping amplitude $J_0$, a large $J_0$ is required to access this regime, see Fig.~\ref{TopoNarrowing}c. 
	
	Here, we note that a topologically trivial chain does not show linewidth narrowing. In a trivial chain, QDs in the lattice act as additional sources of noise for the single photon emission, which, in fact, broadens the linewidth (see SI). This is primarily due to the absence of a topological energy band gap near the emitting state.  
	
	\begin{figure}[h!]
		\centering
		\includegraphics[width=0.48\textwidth]{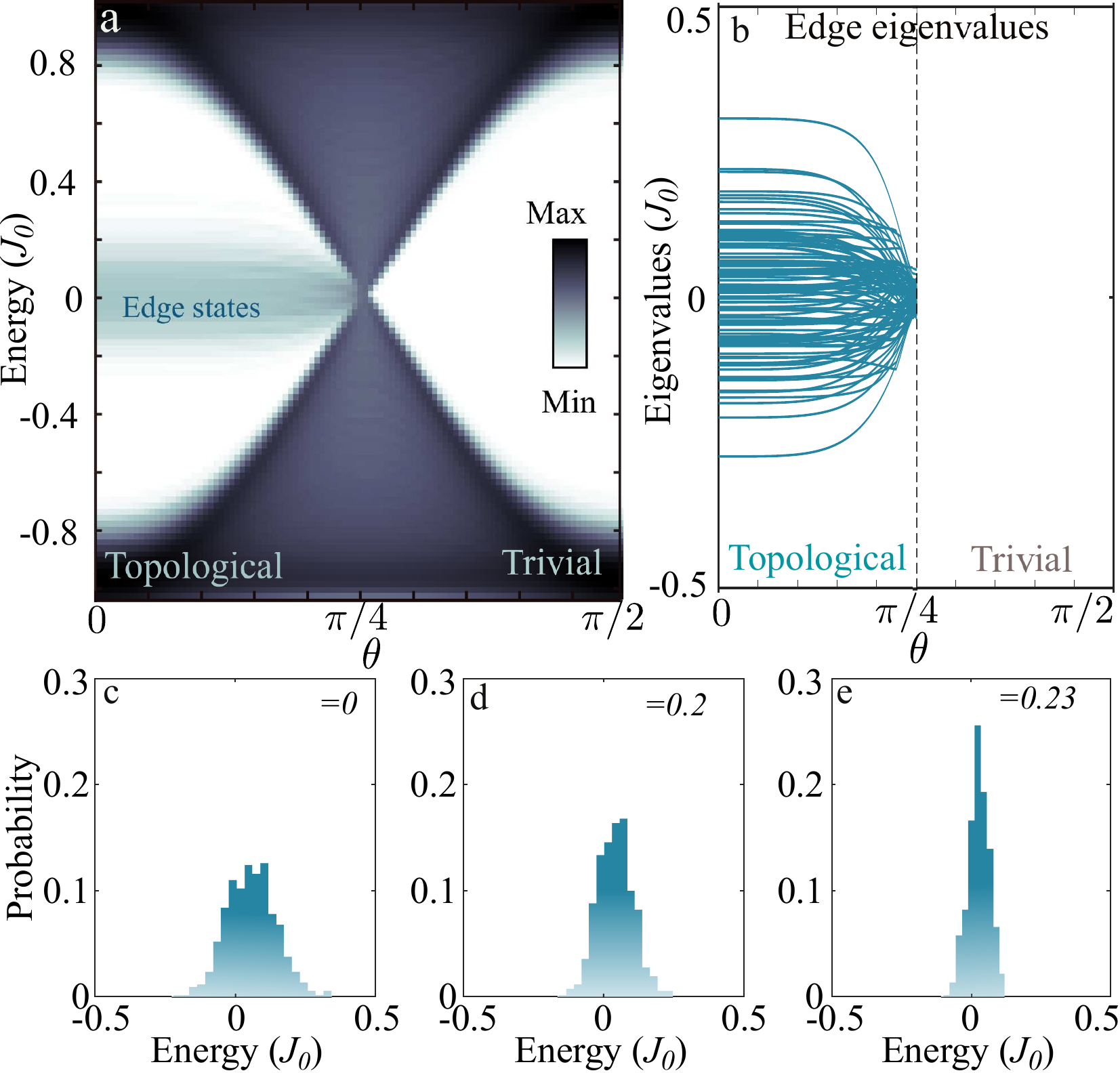}
		\caption{Quasi-static properties of the SSH Hamiltonian. (a) Color plot of the average density of states $D_s(E)=\overline{\text{Tr}[\delta(H-E)]}$ for different $\theta$. (b) Eigenvalues associated with edge states as functions of $\theta$ for different noise realizations (edge states exist for $\theta<\pi/4$). All eigenvalues collapse towards $E=0$ near $\theta=\pi/4$ due to dramatic suppression of fluctuations close to the phase transition. This is the underlying reason for topological linewidth narrowing. (c-e) Probability distributions of edge state eigenvalues for $\theta=0, 0.2\pi$, and $0.23\pi$, respectively. We find that the distribution becomes narrower as $\theta$ approaches $\pi/4$. Here we considered $J_0=5\,meV$ and $\epsilon=0.5\,meV$.}
		\label{DensityOfStates}
	\end{figure}
	
	~\\
	
	To find out the basic mechanism for the topological linewidth narrowing of the single-photon spectrum, we consider an approximate quasi-static picture, where the noise is approximated as a quasi-static change from one static random potential to another. Note that the numerical simulation of a large number of emitters is exceedingly hard due to the exponentially large Hilbert space. However, using the Jordan-Wigner transformation we can convert it to a free Fermionic problem, which is computationally tractable~\cite{LIEB1961407}. In the quasi-static picture, the spectrum is then given by the eigenvalues of an effective single-particle Hamiltonian matrix $H_{jk}=\varepsilon_j\delta_{j,k}-J_{j}\delta_{j+1,k}$. In Fig.~\ref{DensityOfStates}a, we show the density of states $D_s(E)=\overline{\text{Tr}[\delta(H-E)]}$ as a function of energy $E$ and $\theta$. The existence of a finite density of states around $E=0$ and a band gap for $\theta<\pi/4$ evidence the topological non-triviality, whereas the region $\theta>\pi/4$ is topologically trivial. We find that the density of states near energy $E=0$ becomes narrower as $\theta\to \pi/4$. This suggests that here the energy distribution of the edge states is highly constrained. Furthermore, while investigating the eigenvalues of the quasi-static Hamiltonian $H$ in Fig.~\ref{DensityOfStates}b-e, we indeed find that the eigenvalues associated with the edge states show a drastic reduction in the random fluctuation near the transition point $\theta=\pi/4$. This is a manifestation of the topological robustness, which is the origin of the linewidth narrowing observed in Fig.~\ref{TopoNarrowing}. 
	
\begin{figure}[h!]
		\centering
		\includegraphics[width=0.48\textwidth]{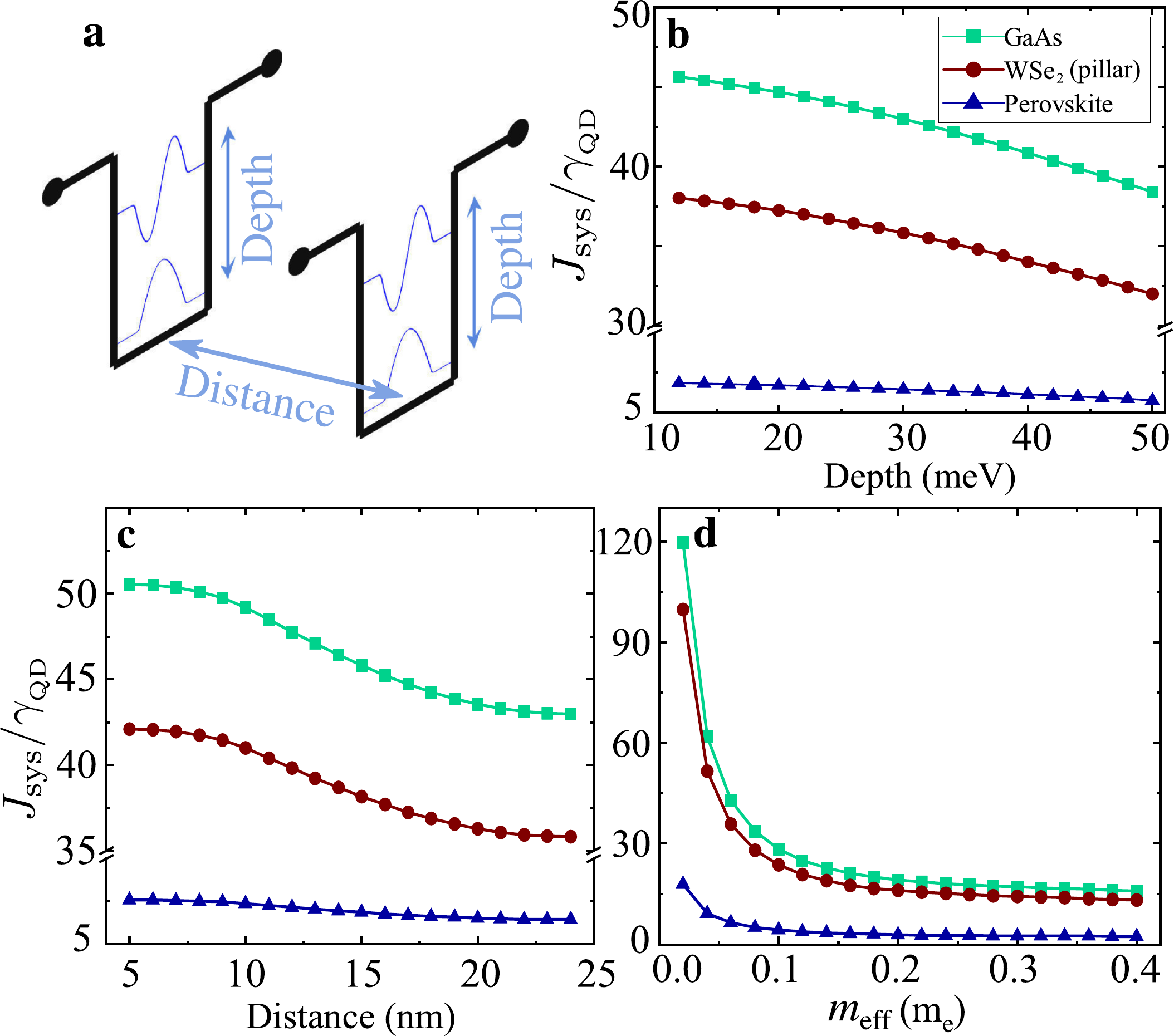}
		\caption{Estimation of the hopping amplitude for different materials. (a) The effective potential energy configuration of two micropillars separated by a distance. (b) Hopping amplitude $J_\text{sys}$ as a function of the energy depth of the micropillars for different materials (GaAs, WSe$_2$, and perovskite). (c) $J_\text{sys}$ as a function of the distance between two micropillars. (d) $J_\text{sys}$ as a function of the effective mass of the micropillar quantum dots. The parameter $\gamma_\text{QD}$ is the linewidth of QDs in respective materials. We consider $m_{\text{eff}}=0.05m_e$, depth $10\,meV$ and distance $5\,nm$, when they are not variables. The single QD linewidth $\gamma_\text{QD}=0.15$ \cite{hours_single_2003}, $0.18$ \cite{palacios-berraquero_large-scale_2017} and $1\, meV$ \cite{raino_single_2016} for GaAs, WSe$_2$, and perovskite, respectively.}
		\label{Fig.4}
	\end{figure}

	\textit{Micropillars:} Micropillars are promising platforms to realize our scheme of topological linewidth narrowing ~\cite{doi:10.1063/1.118135}. Since we require strong quantum hopping ($J$ and $J'$), we estimate the hopping strength for micropillar systems. This system can be described by an effective Hamiltonian $\hat{H}_\text{sys}=-(\hbar^2 \nabla^2)/(2m_\text{eff}) + V(\bf{r})$,
	where the effective mass $m_\text{eff}$ is a material parameter, and the potential energy $V(\bf{r})$ represents the micropillars. A potential well with a certain depth represents a micropillar with a certain height, as schematically shown in Fig.~\ref{Fig.4}a. We investigate how the hopping amplitude $J_\text{sys}$ between two micropillars varies with the effective mass, potential well-depth, and distance between the micropillars. For our simulations, we consider parameters for different materials, such as GaAs, WSe$_2$, and colloidal perovskites quantum dots.
	
	In Fig.~\ref{Fig.4}b, we show the estimation of $J_\text{sys}$ as a function of the potential well-depth for different materials. We find that $J_\text{sys}$ decreases as the depth increases. Similarly, smaller distances between micropillars and smaller effective masses lead to larger $J_\text{sys}$ (see Fig.~\ref{Fig.4}c and d). Note that different materials have different single photon broadening. Thus, it is convenient to compare $J_\text{sys}$ relative to the linewidth $\gamma_\text{QD}$. We find that $J_\text{sys}/\gamma_\text{QD}$ can be as large as $50$, $40$, and $5$ for GaAs, WSe$_2$, and colloidal perovskites quantum dots, respectively. This suggests that our scheme of topological linewidth narrowing within the available materials is achievable.
	
	\begin{figure}[h!]
		\centering
		\includegraphics[width=0.48\textwidth]{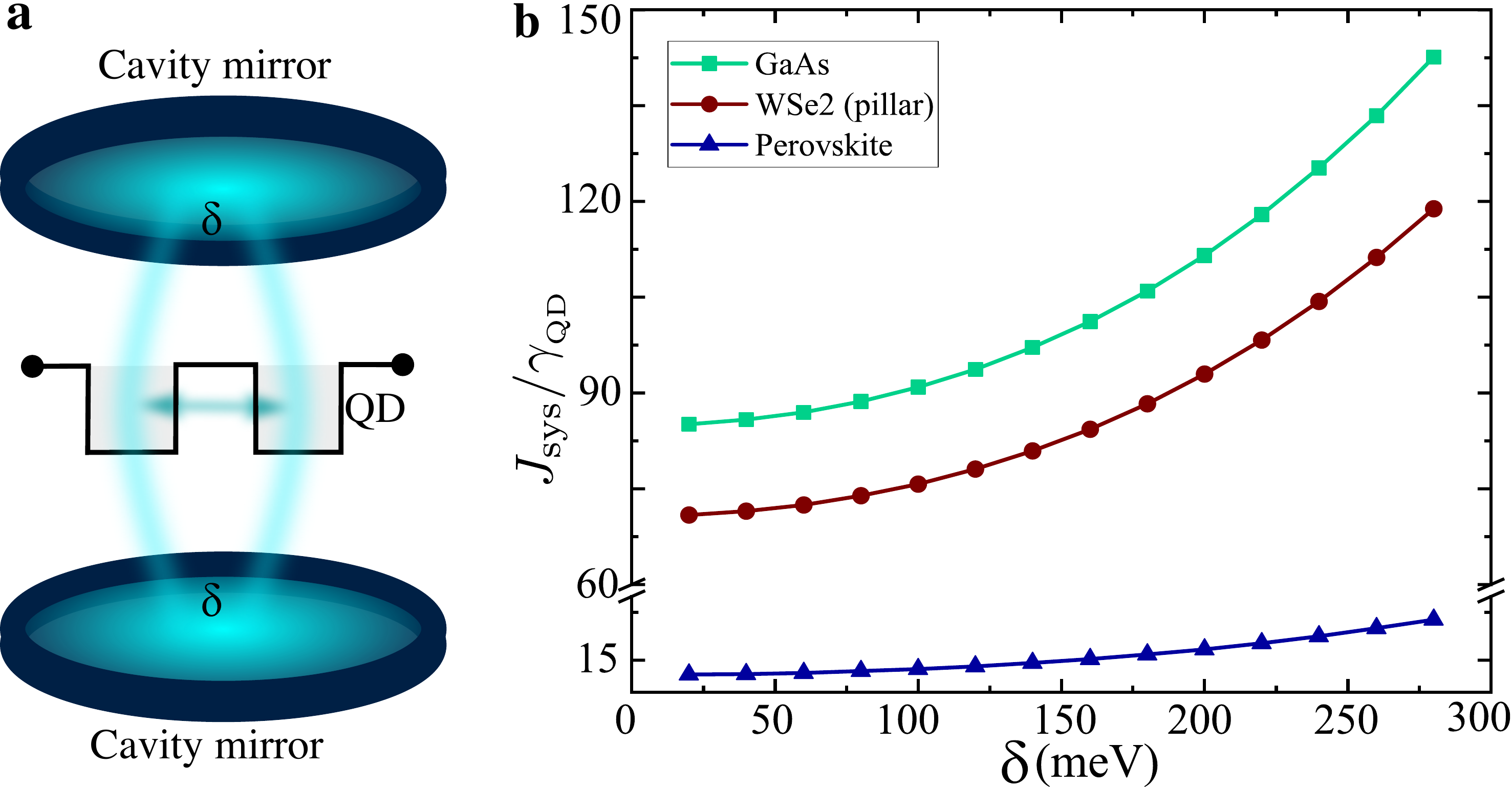}
		\caption{Enhancing the hopping amplitude by light-matter coupling. (a) A schematic representation of the coupling between QDs and an optical cavity. (b) $J_\text{sys}$ as a function of the coupling strength $\delta$ between QDs and the optical cavity for GaAs, WSe$_2$, and Perovskite, respectively. Here, $m_\text{eff} = 0.05m_e$, Depth $=10\,meV$, and $m_c=10^{-4} m_e$.}
		\label{Fig.5}
	\end{figure}
	
	~\\
	
	\textit{Coupling with microcavities:} We find that the effective mass $m_\text{eff}$ is a decisive parameter to control $J_\text{sys}$. However, $m_\text{eff}$ cannot be directly changed for a given material. Here we utilize light-matter coupling in a microcavity system to tune the effective mass. In such a system, QDs are placed inside microcavities such that photons emitted by QDs can coherently couple to the microcavities. The corresponding Hamiltonian is given by
	\begin{equation}
		H_\text{tot}=\begin{pmatrix}
			H_\text{sys} & \delta\\
			\delta & -(\hbar\nabla)^2/(2m_{c})
		\end{pmatrix},
		\label{EqQD-MC}
	\end{equation}
	where $m_c$ is the effective mass of the cavity photons ($m_c \ll m_\text{eff}$), $\delta$ determines the coupling strength between cavity photons and QDs. Since QDs coherently couple to the cavity, it opens up an additional channel for the hopping process. This can be interpreted as a reduction in the effective mass, and hence an enhancement in $J_\text{sys}$ is obtained (see Fig.~\ref{Fig.5}).
	
	~\\
	
	\textit{\textbf{Discussion:}} Our theoretical model and the scheme of topological linewidth narrowing are promisingly extended to any two-level quantum system, e.g., superconducting qubits, NV centers, and trapped atoms or ions. Moreover, it can be applied to enhance the operating temperature of quantum emitters since the topological linewidth narrowing in quantum SSH chains can effectively compensate for thermal broadening at high temperatures. In fact, since the linewidth can be arbitrarily reduced ($\gamma_\text{SSH}\to 0$) by increasing the size of the SSH chain ($N\to\infty$), the operating temperature can be, in principle, enhanced to an arbitrarily high value. The experimental realization of an SSH chain with QDs requires positioning them in regular patterns, which was a long-standing challenge. However, thanks to the recent progress in micro-nanofabrication, one can achieve deterministic arrays of quantum emitters~\cite{branny2017deterministic}. For this, one can first define an SSH lattice with standard lithography and subsequently guide the growth, transfer, or filling of QDs onto the lattice, as discussed in detail in the Supplementary Information \cite{palacios-berraquero_large-scale_2017, tran_deterministic_2017, luo2018deterministic, marago2013optical, fulmes2015self, kohmoto1999site, jons2013triggered}. Another challenge could be to selectively channel emitted photons to the detector only from the edge site, i.e., $j=1$. This could be solved by coupling a waveguide structure between the edge site and the detector \cite{PhysRevX.2.011014, javadi_single-photon_2015}. 
	
	~\\
	
	\textit{\textbf{Conclusion:}} We have developed a scheme for the linewidth narrowing of single photon emission by arranging quantum emitters in a topological non-trivial SSH lattice. We observed that near the topological phase transition point, the linewidth dramatically decreases to a small value compared to that of an isolated emitter. The linewidth could be arbitrarily small near the phase transition point for a large system size. However, a strong hopping amplitude is required to maintain a sizeable band gap. We find that the linewidth narrowing is microscopically linked to the suppression of eigenvalue fluctuations near the transition point. We have provided analyses of promising physical systems for experimental realization. Our scheme can be integrated into photonic chip-based quantum computers or information processors. Since our scheme does not require external control, it is scalable, so it can ultimately be used in large numbers in photonic platforms.
	
	~\\
	
	\textbf{Acknowledgments}
	
	Q. X. gratefully acknowledges funding support from the National Natural Science Foundation of China (Grant no. 12020101003, and 12250710126) and strong support from the State Key Laboratory of Low-Dimensional Quantum Physics at Tsinghua University. S. G. gratefully acknowledges funding support from the Excellent Young Scientists Fund Program (Overseas) of China, the National Natural Science Foundation of China (Grant No. 12274034) and the start up grant from Beijing Academy of Quantum Information Sciences.

	\bibliography{reference}
	
	\newpage
	
	\section{Supplymentary Materials}
	
	\subsection{Noise Model}
	The noise we considered in our system is in the form of a time-dependent on-site energy fluctuation; $\varepsilon_j(t)=\epsilon\delta\varepsilon_j(t)$, where $\varepsilon$ indicates the strength of the fluctuation. In one iteration, the noise along different time steps is shown in Fig.~\ref{Fig.2}(a), and the probability distribution can be seen in Fig.~\ref{Fig.2}(b). In the case of only one quantum emitter, ideally, it should emit photons at only one energy ($\hbar\omega=0$). However, when we include the noise $\varepsilon_j(t)$, the emission spectrum broadens as shown by Fig.~\ref{Fig.2}(c)). The numerical data fits well with the Lorentzian function obtained with our analytical theory based on the retarded Green's function. 
	
	\renewcommand*{\thefigure}{S1}
	\begin{figure}[h!]
		\centering
		\includegraphics[width=0.9\textwidth]{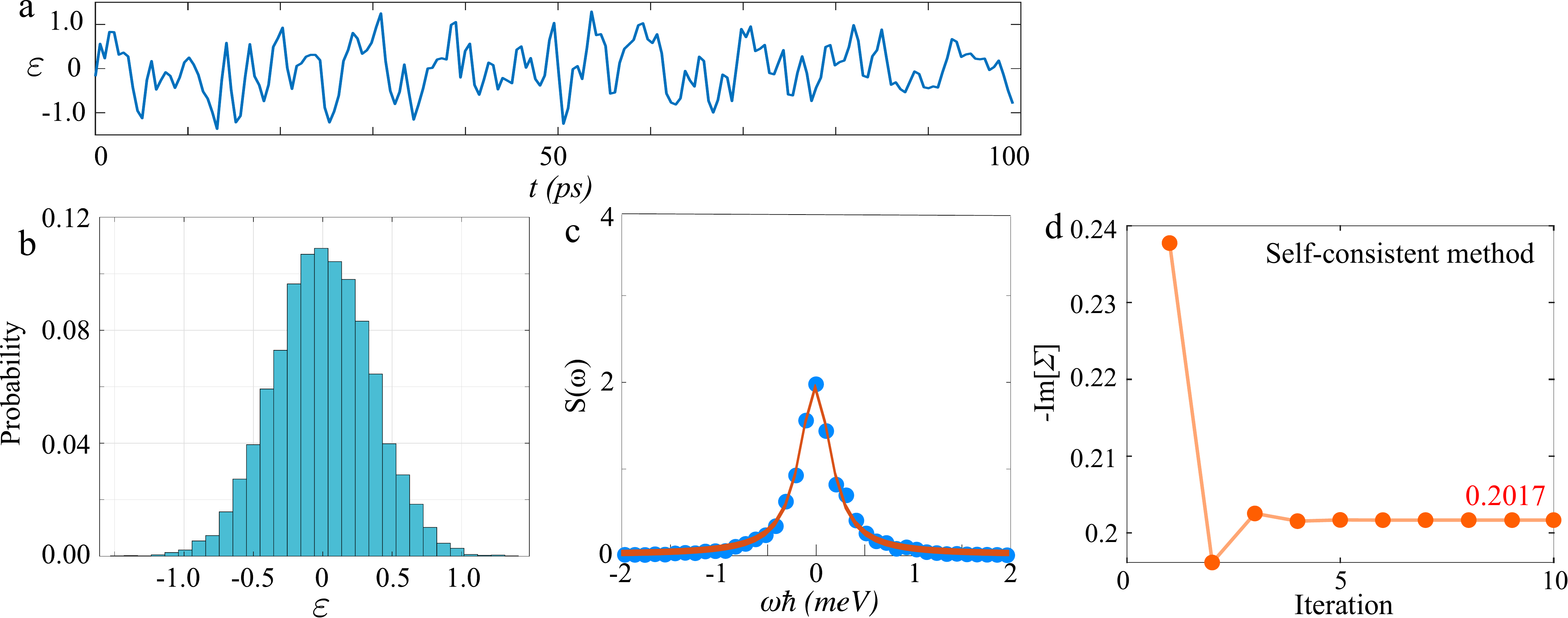}
		\caption{The noise model and related properties. (a) Example of a Gaussian correlated noise realization as a function of time. (b) Distribution of the Gaussian random variable $\delta\epsilon_j$ for a lattice site $j$. (c) Numerically calculated emission spectrum $S(\omega)$ of a single QD with strong time-dependent noise. Solid circles are numerical data and the solid red line is a Lorentzian fit with a full width at half maximum $\gamma_\text{QD}=0.4\, meV$. (d) The negative imaginary part of the self-energy $-\Im [\Sigma]$ with the self-consistent iteration steps. The convergence found at $-\Im [\Sigma]=0.2017\, meV$ which corresponds to $\gamma_\text{QD}=0.403\, meV$. Here we consider $\epsilon=0.5$ and $\tau=0.5$ ps. The spectrum is averaged over $10$ noise realizations.}
		\label{Fig.2}
	\end{figure}

	\subsection{Green's Function, Self-Energy, and Spectrum}
	We consider a Gaussian correlated noise with the correlation function: $F(t-t^\prime)=\overline{\delta\varepsilon_j(t)\delta\varepsilon_j(t^\prime)}=\epsilon^2e^{-(t-t^\prime)^2/(2\tau^2)}$, where the over-line represents the average over different noise realizations. The parameter $\tau$ represents the correlation time of the noise. It is convenient to obtain the Fourier transform of the correlation function
	\begin{equation}
		\widetilde{F}(\omega)=\int_{-\infty}^\infty dt\, F(t)e^{-i\omega t}=\epsilon^2\tau\sqrt{2\pi}e^{-(\omega\tau)^2/2}
	\end{equation}
	In the case of a single emitter, it is equivalent to a zero-dimensional Fermion. The retarded Green's function of the single emitter without noise is given by $G^R(\omega)=\lim_{\eta\to 0}(\hbar\omega + i\eta)^{-1}$. With noise, the average Green's function can be written as
	\begin{equation}
		\overline{G}^R(\omega)=\lim_{\eta\rightarrow0}\frac{1}{\hbar\omega+i\eta-\Sigma}
		\label{AvgGreen}
	\end{equation}
	where $\Sigma$ is the self-energy of the emitter due to the
	fluctuating noise. The spontaneous emission spectrum is given by
	\begin{equation}
		S(\omega)=-\frac{1}{\pi}\Im\overline{G}^R(\omega)
	\end{equation}
	which is a Lorentzian distribution function with the full width at half maximum $\gamma_\text{QD} = -2\Im\Sigma$. In the lowest order of the Dyson series, the self-energy is given by
	\begin{equation}
		\Sigma  \approx\int_{-\infty}^\infty\frac{d\omega}{2\pi}\widetilde{F}(\omega)G^R(\omega)=-\frac{i\epsilon^2\tau\sqrt{\pi}}{\hbar\sqrt{2}}
	\end{equation}
	Thus in the lowest order: $\gamma_\text{QD} = -2\Im\Sigma=(\epsilon^2\tau\sqrt{2\pi})/\hbar$. For $\epsilon = 0.5\, meV$ and $\tau=0.5\,ps$, we find $\gamma_\text{QD} \approx 0.47\, meV$. A more accurate analytical estimate can be obtained with the self-consistent Dyson equation, which implies
	\begin{equation}
		\Sigma = \int_{-\infty}^\infty\frac{d\omega}{2\pi}\widetilde{F}(\omega)\overline{G}^R(\omega)
		\label{Self-consistent}
	\end{equation}
	Since $\overline{G}^R(\omega)$ itself depends on $\Sigma$, both sides of Eq.~\ref{Self-consistent} involve $\Sigma$. Thus, a solution for $\Sigma$ can be obtained with the self-consistent method. In this method, we start with a guess for the value of $\Sigma$ to obtain the right-hand-side of Eq.~\ref{Self-consistent}, which gives us a new value for $\Sigma$. Then we recalculate the right-hand side with the new $\Sigma$ and repeat the process until the convergence for the value of $\Sigma$ is reached. Here, we do it numerically, considering an initial value $-i\times 10^{-5}\, meV$, and after $10$ iterations, we obtain a convergence at $\Sigma = -0.2017i\, meV$. In this self-consistent method, the width of the spectrum is thus $\gamma_\text{QD} = 0.403\, meV$ (see Fig. \ref{Fig.2}d) which is almost the same as the exact numerical value $0.40\, meV$(see Fig.~\ref{Fig.2}c).

	\vspace{0.5cm}
	\hspace{-0.35cm}{\large \textbf{Proof of the relation between $G^R(\omega)$ and $S(\omega)$:}}
	Here, in this section, we prove the relation between the spectrum and the retarded Green's function for a single QD: $S(\omega) = -\Im G^R(\omega)/\pi$. The Hamiltonian of a QD with constant energy $\varepsilon$ is given by
	\begin{equation}
		\begin{aligned}
			\mathcal{H}=\varepsilon \sigma^+\sigma^- 
		\end{aligned}
	\end{equation}
	The time evolution for an operator is given by
	\begin{equation}
		\begin{aligned}
			\sigma^+(t) = e^{i\mathcal{H}t/\hbar} \sigma^+ e^{-i\mathcal{H}t/\hbar}\\
		\end{aligned}
	\end{equation}
	The correlation function is then given by
	\begin{equation}
		\begin{aligned}
			C(t,t') = \langle 1 | \sigma^+(t) \sigma^-(t') |1 \rangle =  e^{i\varepsilon t/\hbar} \langle 1 | \sigma^+ e^{-i\mathcal{H}(t-t')/\hbar} \sigma^- |1\rangle e^{-i\varepsilon t'/\hbar} = e^{-i\varepsilon (t'-t)/\hbar}
		\end{aligned}
	\end{equation}
	where $\ket{1}$ is the excited state of the QD. Let us define Green's functions as
	\begin{equation}
		\begin{aligned}
			G^R(\omega) = \lim_{\eta\to 0+} \frac{1}{\hbar\omega-\varepsilon + i\eta} = -i  \lim_{\eta\to 0+} \int_0^\infty dt\, e^{i\omega t -i\varepsilon t/\hbar -\eta t/\hbar} \\
			G^A(\omega) = \lim_{\eta\to 0+} \frac{1}{\hbar\omega-\varepsilon - i\eta} = i \lim_{\eta\to 0+}\int_0^\infty dt\, e^{-i\omega t + i\varepsilon t/\hbar -\eta t/\hbar}
		\end{aligned}
	\end{equation}
	where $G^R$ and $G^A$ are the retarded and advanced Green's functions respectively. Note that the two Green's functions satisfy,
	\begin{equation}
		\begin{aligned}
			\Im[G^R(\omega)] = \frac{-i[G^R(\omega) - G^A(\omega)]}{2} =  \lim_{\eta\to 0+}  \left[ -\eta G^R(\omega) G^A(\omega) \right]
		\end{aligned}
	\end{equation}
	Finally, let us rewrite the correlation function as
	\begin{equation}
		\begin{aligned}
			C(t,t') = \lim_{\eta\to 0+} e^{-i\varepsilon t'/\hbar -\eta t/\hbar} e^{i\varepsilon t/\hbar -\eta t/\hbar}
		\end{aligned}
	\end{equation}
	The spectrum is then given by
	\begin{equation}
		\begin{aligned}
			S(\omega) &= \frac{\int_0^\infty dt\int_0^\infty dt' \,C(t,t')^\prime e^{-i\omega(t-t^\prime)} }{2\pi\int_0^\infty dt\, C(t,t)} 
			= \lim_{\eta\to 0+} \frac{G^R(\omega)G^A(\omega)}{2\pi \int_0^\infty dt\, e^{-2\eta t}} \\
			&= \lim_{\eta \to 0+} \frac{\eta G^R(\omega)G^A(\omega)}{\pi} 
			= -\frac{1}{\pi}\Im{G}^R(\omega)
		\end{aligned}
	\end{equation}
	Here, we have considered the QD without noise. In the presence of noise, the exact solution would depend on the exact noise realization, which follows a random distribution. In general, the exact analytical solution is impractical to obtain for each random noise realization. So instead, we analyze the average properties of the system. The system can be described in this average picture by an average Green's function $\overline{G}^R(\omega)$. This average Green's function can be obtained using perturbation theory with Dyson series expansion, as we have used previously in Eq.~\ref{AvgGreen}.

	\subsection{The Su-Schrieffer-Heeger Model}
	The Su-Schrieffer-Heeger (SSH) model describes particles hopping on a one-dimensional chain lattice with staggered hopping strength. The Hamiltonian of an SSH chain consisting of $N=2M$ sites can be expressed as
	\begin{equation}
		\hat H=J \sum_{j=1}^M(\hat a_{j}^\dagger \hat b_{j} + h.c.) + J'\sum_{j=1}^{M-1}(\hat b_{j}^\dagger \hat a_{j+1} + h.c.)
		\label{S5}
	\end{equation}
	where $J(J')$ represents the hopping amplitudes between different sites, $\hat a_j$ ($\hat a_j^\dagger$) and $\hat b_j$ ($\hat b_j^\dagger$) is the annihilation (creation) operators for odd and even site numbers. In our scheme, we consider $J$ and $J'$ are positive and real-valued. We use the Fourier transform
	\begin{equation}
		\begin{aligned}
			\hat a_k = \frac{1}{\sqrt{M}}\sum_{j=1}^M \hat a_j e^{-ikr_j} \\
			\hat b_k = \frac{1}{\sqrt{M}}\sum_{j=1}^M \hat b_j e^{-ikr_j}
		\end{aligned}
	\end{equation}
	and considering periodic boundary condition, Eq.~\ref{S5} can be rewritten as
	\begin{equation}
		\begin{aligned}
			\hat H=& \sum_{k}[ J(\hat a_k^\dagger \hat b_{k} + b_k^\dagger \hat a_{k}) + J'(\hat a_{k}^\dagger \hat b_{k} e^{-ik}+ b_{k}^\dagger \hat a_{k} e^{ik}) ] \\
			=&\sum_k [ (J+J'e^{-ik})\hat a_k^\dagger\hat b_k + (J+J'e^{ik})\hat b_k^\dagger\hat a_k ] \\
			=&\sum_k \psi_k^\dagger H_k \psi_k
		\end{aligned}
	\end{equation} 
	where the two-band Hamiltonian matrix $H_k$ in the reciprocal space is given by
	\begin{equation} H_k = 
		\begin{pmatrix}
			0 & J + J'e^{-ik} \\
			J + J'e^{ik} & 0 
		\end{pmatrix}
	\end{equation}
	and the two-band wave function $\psi_k$ is given by
	\begin{equation} \psi_k=
		\begin{pmatrix}
			\hat a_k\\
			\hat b_k 
		\end{pmatrix}
	\end{equation}
	The corresponding dispersion relation for the bulk states is given by
	\begin{equation}
		\begin{aligned}
			E(k)&=\pm \left|J+J^\prime e^{-ik}\right|\\
			&=\pm \sqrt{J^2+{J^\prime}^2+2JJ^\prime \cos k}.
		\end{aligned}
	\end{equation}
	The dispersion relation changes the qualitative character depending on the choice of the hopping amplitudes $J$ and $J'$ (see Fig. \ref{Winding}). As can be seen, for $J\neq J^\prime$ the dispersion shows an energy gap opening. The two-band Hamiltonian matrix $H_k$ can be written as
	\begin{equation}
		\begin{aligned}
			H_k&=f_0(k)\hat \sigma_0+f_x(k)\hat \sigma_x+f_y(k)\hat \sigma_y+f_z(k)\hat \sigma_z\\
			&=f_0(k)\hat \sigma_0+\boldsymbol{f}(k)\cdot\boldsymbol{\hat \sigma}
		\end{aligned}
	\end{equation}
	where $\hat\sigma_i$ are the Pauli matrices. For our considered SSH model, $f_0(k)=0,\, f_x(k)=J+J^\prime \cos k,\, f_y(k)=J^\prime \sin k$, and $f_z(k)=0$.
	
	\renewcommand*{\thefigure}{S2}
	\begin{figure}[h!]
		\centering
		\includegraphics[width=0.6\textwidth]{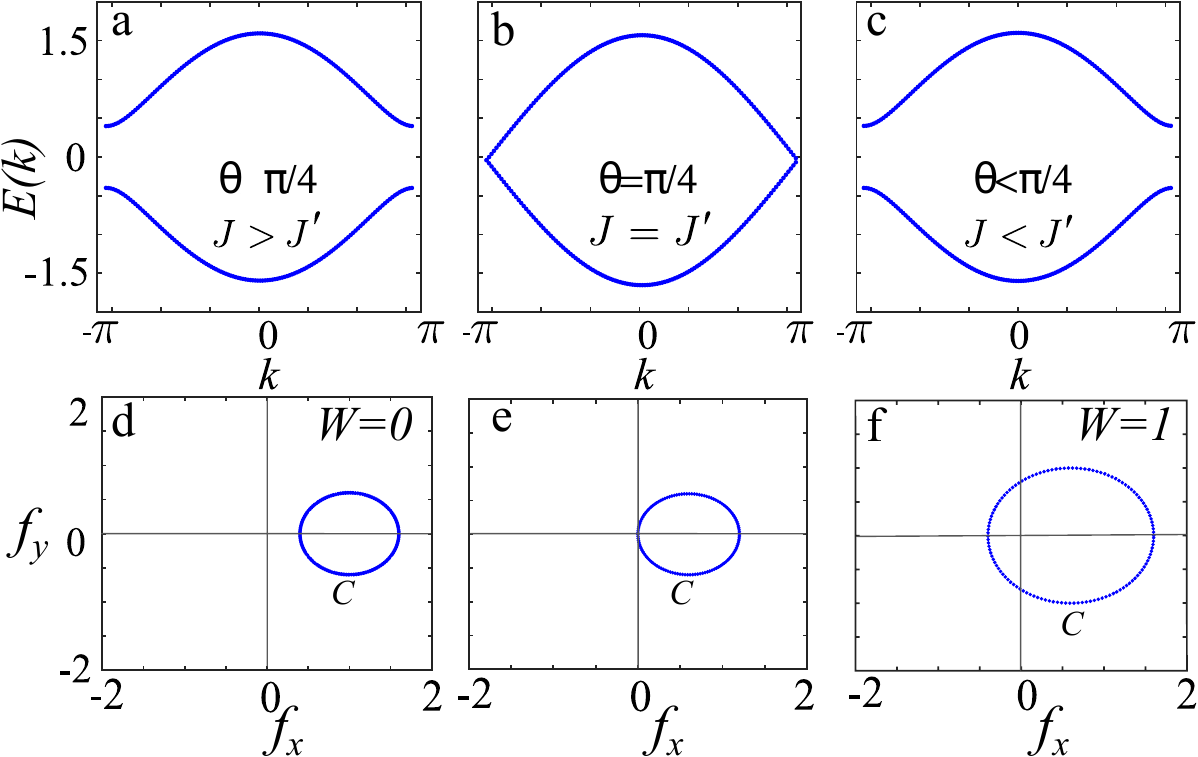}
		\caption{Energy band structures and the winding plots of the SSH model for different choices of hopping amplitudes. The parameters are $J=1$, $J'=0.6$ for (a,d), $J=0.6$, $J'=0.6$ for (b,e) and $J=0.6$, $J'=1$ for (c,f).}
		\label{Winding}
	\end{figure} 
	
	The nontrivial nature of the SSH Hamiltonian can be quantified with an integer called the winding number $W$
	\begin{equation}
		\begin{aligned}
			W&=\frac{1}{2\pi}\int^\pi_{-\pi}\left[\boldsymbol{\tilde{f}}(k)\times\frac{d}{dk}\boldsymbol{\tilde{f}}(k)\right]_zdk\\
			&\equiv \frac{1}{2\pi i}\int_{-\pi}^{\pi}dk\frac{d}{dk}log[h(k)]
		\end{aligned}
	\end{equation}
	where $\boldsymbol{\tilde{f}}=\boldsymbol{f}/|\boldsymbol{f}|$, and $h(k)=J+J^\prime e^{ik}$. The winding number has a fascinating geometric interpretation on the plane spanned by the vector $\boldsymbol{f}(k)$. The endpoint of the vector $\boldsymbol{f}(k)$ traces out a closed circular path $C$ on the $f_x$-$f_y$ plane as the wavenumber $k$ runs through the Brillouin zone ($-\pi$ to $\pi$). This circular path has the radius $J^\prime$ and is centered at $(J,\,0)$. The winding number $W$ counts the number of times the loop winds around the origin of the $f_x$-$f_y$ plane.
	
	In Fig.~\ref{Winding}(d), as we can see, when the path $C$ does not enclose the origin of the $f_x$-$f_y$ plane (for $J> J'$), the system has a winding number $W=0$, which is referred to as the topologically trivial phase. If the path $C$ encloses a single complete loop around the origin (for $J<J'$), the system is topologically nontrivial with winding number $W=1$, as shown in Fig.~\ref{Winding}f. Note that $J=J'$ corresponds to the phase transition point, at which the path $C$ crosses the origin with a winding number undefined (see Fig.~\ref{Winding}(e)). 
	
	Note that the topological character of the band structure depends on the relative value of $J$ with respect to $J'$. Thus, it is convenient to parameterize them with a single variable $\theta$
	\begin{equation}
		\begin{aligned}
			J = J_0 \sin^2\theta \\
			J' = J_0 \cos^2\theta
		\end{aligned}
	\end{equation}
	where $J_0$ is a constant. Here, the topological trivial and nontrivial phases correspond to $\theta>\pi/4$ (Fig.~\ref{Winding}(d)) and $\theta<\pi/4$ (Fig.~\ref{Winding}(g)) respectively, and the phase transition point is given by $\theta=\pi/4$ (Fig.~\ref{Winding}(f)).


	\subsection{Micropillar Quantum Dots}
	
	\renewcommand*{\thefigure}{S3}
	\begin{figure*}[t]
		\centering
		\includegraphics[width=0.8\textwidth]{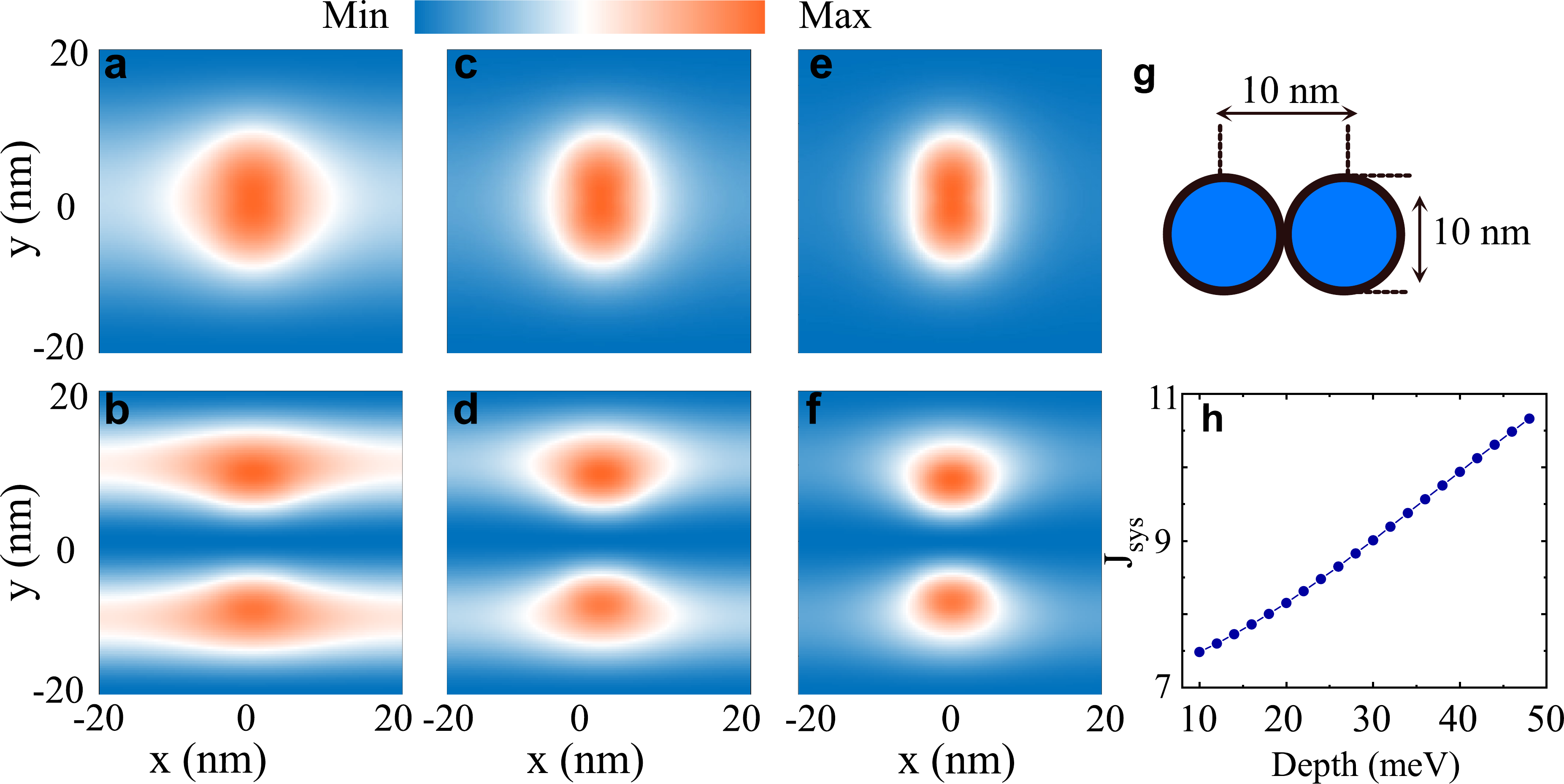}
		\caption{Intensity distribution of the ground state wavefunction ($\psi_g$) and the first excited state wavefunction ($\psi_e$) of the micropillar systems with different heights. (\textbf{a}) and (\textbf{b}) are $|\psi_g|^2$ and $|\psi_e|^2$ for height 10 $meV$,  (\textbf{c}) and (\textbf{d}) are $|\psi_g|^2$ and $|\psi_e|^2$ for height 18 $meV$, (\textbf{e}) and (\textbf{f}) are $|\psi_g|^2$ and $|\psi_e|^2$ for height 28 $meV$. (\textbf{g}) A schematic for two micropillars with a diameter of 10 $nm$ and separated 10 $nm$. (h) Hopping strength $J_{\text{sys}}$ as a function of the height of micropillars.}
		\label{S3}
	\end{figure*}
	
	\renewcommand*{\thefigure}{S4}
	\begin{figure*}[t]
		\centering
		\includegraphics[width=0.8\textwidth]{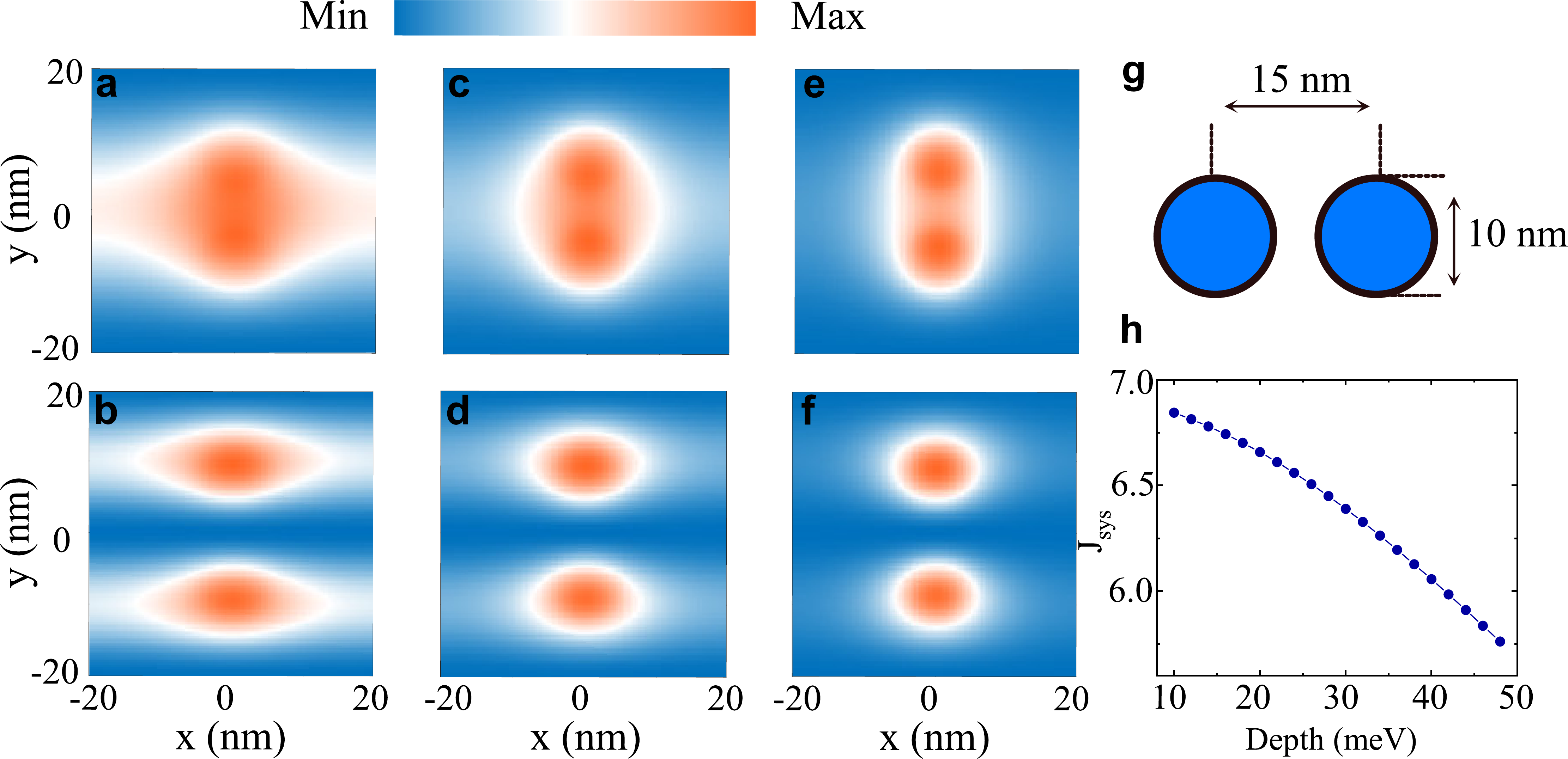}
		\caption{Intensity distribution of the ground state wavefunction ($\psi_g$) and the first excited state wavefunction ($\psi_e$) of the micropillar systems with different heights. (\textbf{a}) and (\textbf{b}) are $|\psi_g|^2$ and $|\psi_e|^2$ for height 10 $meV$,  (\textbf{c}) and (\textbf{d}) are $|\psi_g|^2$ and $|\psi_e|^2$ for height 18 $meV$, (\textbf{e}) and (\textbf{f}) are $|\psi_g|^2$ and $|\psi_e|^2$ for height 28 $meV$. (\textbf{g}) A schematic for two micropillars with a diameter of 10 $nm$ and separated 15 $nm$. (h) Hopping strength $J_{\text{sys}}$ as a function of the height of micropillars.}
		\label{S2}
	\end{figure*}
	
	In the main text, we consider semiconductor micropillars as promising physical systems to realize our scheme. We have calculate the hopping amplitude between two micropillars, which provides an estimation for $J$ and $J'$ of the SSH chain. While estimating the hopping amplitudes, we find that the influence of the micropillar heights on the hopping strength can be different depending on the displacement between the micropillars. We show in the main text that the hopping strength between two QDs decreases with the increasing height of the micropillars. Note that the hopping strength can be calculated from the energy difference between the lowest two energy levels of two micropillars, which comes from the hybridization between the ground states of the individual micropillars. Here, we show the intensity distribution of the ground state $\psi_g$ and first excited states $\psi_e$ with different micropillar heights in Fig.~\ref{S2}(a) to (f). In this case, we consider the distance between two micropillars is $15$ $nm$, as shown in Fig.~\ref{S2}(g). The hopping strength between two QDs decreases with increasing the heights of the micropillars (see Fig.~\ref{S2}(h)). 
	
	However, we find that when the distance between the two micropillars is smaller, for example, $10$ $nm$, as shown in Fig.~\ref{S3}(g). In this case, we show the intensity distributions of $\psi_g$ and $\psi_e$ for different micropillar heights in Fig.~\ref{S3}(a) to (f) as well. But when we increase the heights of the micropillars, the hopping strength between the two QDs increases, as shown in Fig.~\ref{S3}(h).
	
	\subsection{From Quantum Dots to Free Fermions}
	A system involving many QDs is, in general, a computationally hard problem. This is due to the exponentially large Hilbert space of the system. For $N$ QDs, the Hilbert space size is $2^N$, analyzing which is beyond our computational capability. However, for our QD SSH chain, we use the Jordan-Wigner transformation to simplify the computational problem. The Jordan-Wigner transformation is given by
	\begin{equation}
		\begin{aligned}
			\sigma_j^+=u_j^\dag c_j^\dag\\
			\sigma_j^-=u_j c_j
		\end{aligned}
	\end{equation}
	where $u_j = \exp[i\pi\sum_{k=1}^{j-1}c_k^\dag c_k]$, and $c_k (c_k^\dag)$ are the Fermionic annihilation (creation) operators. Using this transformation, we find that the
	Hamiltonian $H$ is equivalent to a fermionic Hamiltonian
	\begin{equation}
		\begin{aligned}   
			\mathcal{H}_F&=\sum_j\varepsilon_jc_j^\dag c_j-\sum_jJ_j(c_j^\dag c_{j+1}^\dag+c_{j+1}c_j)\\
			&=\sum_{jk}c_j^\dag H_{jk}c_k
		\end{aligned}
	\end{equation}
	This Hamiltonian represents an SSH chain of free Fermions. Here, the matrix $H$ is the single particle Hamiltonian written in position space, such that $H_{jk}=\varepsilon_j\delta_{j,k}-J_{j}\delta_{j+1,k}$. This single-particle Hamiltonian can be efficiently diagonalized.  Let us now introduce the time-dependent annihilation (creation) operators $c_j(t)$ (or $c_j^\dag(t)$)
	\begin{equation}
		c_j(t)=U^\dag(t)c_jU(t)
	\end{equation}
	where $U(t)$ is the unitary operator describing the time evolution of the system. For a time-independent Hamiltonian $\mathcal{H}_F=\mathcal{H}_F(t=0)$,  the unitary operator $U(t)=exp[-i\mathcal{H}_F(t=0)t/\hbar]$. However, in the presence of noise the Hamiltonian $\mathcal{H}$ (equivalently $\mathcal{H}_F$) is time-dependent due to the time dependence of $\varepsilon_j(t)$. It is thus convenient to discretize
	the time $t$ into intervals of $\Delta t$, such that the time runs as $t_m=m\Delta t$ where $m$ is an integer between 0 and $M$ and $\Delta t\to 0$. Then the unitary evolution operator at a time $t= M\Delta t$ is given by
	\begin{equation}
		U(t)=\lim_{\Delta t\to 0}\prod_{m=1}^Me^{-im\Delta t \mathcal{H}_F(t_m)/\hbar}
	\end{equation}
	We are interested in the emission from the $j^\text{th}$ site. The corresponding Jordan-Wigner transformation is given by $\sigma_j^+=u_j^\dag c_j^\dag$ and $\sigma_j^-=u_j c_j$. This implies
	\begin{equation}
		\begin{aligned}   
			c_j(t)&=U^\dag(t)u_j^\dag \sigma_j^- U(t) \\
			c_j^\dag(t)&=U^\dag(t) u_j \sigma^+_jU(t)
		\end{aligned}
	\end{equation}
	Furthermore, the correlation function for the site $j$ is given by
	\begin{equation}
		\braket{\sigma_j^+(t)\sigma_j^-(t^\prime)}=\braket{c_j^\dag(t)c_j(t^\prime)}
	\end{equation}
	We have thus proved that the local unequal time correlation function at a site $j$ is given by the local correlation in an equivalent site in a free Fermion theory. Thus we find
	\begin{equation}
		C_j(t,t^\prime)=\braket{c_j^\dag(t)c_j(t^\prime)}
	\end{equation}
	Since the Hamiltonian $\mathcal{H}_F$ represents free Fermions, the correlation function $C_j(t,t^\prime)$ is determined by the single particle Hamiltonian $H$. The Heisenberg equation for the operator $c_j(t)$ is given by
	\begin{equation}
		\dot{c}_j(t)=\frac{i}{\hbar}[\mathcal{H}_F, c_j(t)]
	\end{equation}
	Using the Fermionic anticommutation relations, we find that the time-dependent operator $c_j(t)$ obeys Schrodinger's equation
	\begin{equation}
		\dot c_j(t)=-\frac{i}{\hbar}\sum_k H_{jk}c_k(t)
	\end{equation}
	This reinstates the fact that $\mathcal{H}_F$ represents free Fermions with the single particle Hamiltonian $H$.

	\subsection{Topologically Trivial Lattice}
	
	\renewcommand*{\thefigure}{S5}
	\begin{figure*}[t]
		\centering
		\includegraphics[width=0.8\textwidth]{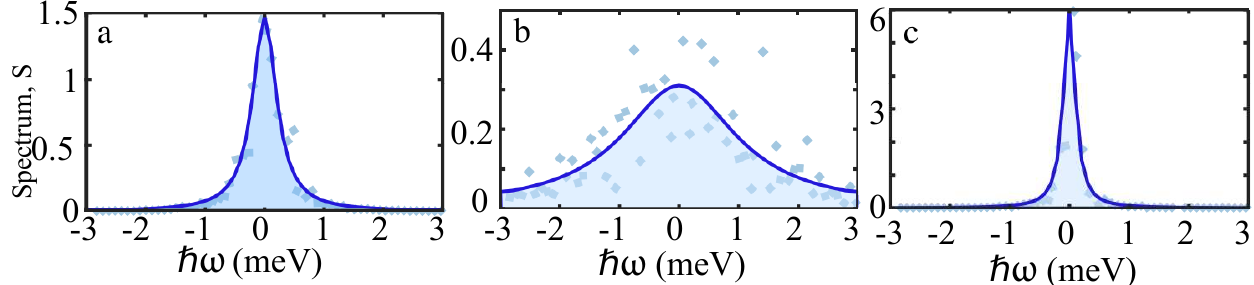}
		\caption{Comparison with trivial chains. (a-c), Single photon spectra of a QD, and topologically trivial and nontrivial chains respectively. The spectrum from the topologically trivial chain has a larger width than an isolated QD.}
		\label{3spectrums}
	\end{figure*}
	
	We here consider a topologically trivial lattice of QDs. The corresponding Hamiltonian of the system is given by
	\begin{equation}
		\mathcal{H} = \sum_j\varepsilon_j(t)\sigma_j^+\sigma_j^- + \frac{J_0}{2} \sum_j (\sigma_j^+\sigma_{j+1}^-+\sigma_{j+1}^+\sigma_j^-)
	\end{equation}
	We calculate the corresponding single photon spectrum, as shown in Fig.~\ref{3spectrums}(b). Compared to a single QD (Fig.~\ref{3spectrums}(a)), the single photon spectrum of a trivial chain becomes wider. This is due to the absence of a quantum state inside a band gap for a trivial lattice. Thus a large number of bulk states can interact with any other state. In this way, these states act as an additional source of noise to the single photon emission.

	\subsection{Perspective on the Experimental Realization}
	
	In practice, nanomaterials that exhibit photon emission with high brightness and indistinguishability mainly include (\textbf{a}) traditional III-V semiconductor quantum dots grown with epitaxy, including GaAs/InAs, GaAs/AlGaAs, GaN/AlGaN, $etc$. (\textbf{b})defects formed, or strain induced SPE in 2D transition metal dichalcogenides, hexagonal boron nitride, $etc$ (\textbf{c})solution-processed colloidal nanocrystal with 0D character, including CdSe/ZnS, colloidal perovskite, $etc$.
	In the main text, we use parameters of GaAs, WSe$_2$, and colloidal perovskite quantum dots (TABLE S1) obtained from experiments to numerically study the linewidth narrowing effect in an SSH chain. The sizes of these quantum dots are typically on a scale of 10 to 100 nm. They have been extensively studied in experiments, with mature fabrication or synthesis procedures and also on-demand spectral properties. Therefore, they are promising for demonstrating the effectiveness of topologically narrowing the linewidth of quantum dot emission.

	\renewcommand*{\thetable}{S1}
	\begin{table}[h]
		\scriptsize
		\newcommand{\tabincell}[2]{\begin{tabular}{@{}#1@{}}#2\end{tabular}}
		\renewcommand{\arraystretch}{0.75}
		\centering
		\begin{tabular}{|c|c|c|c|c|c|c|c|c|}
			\hline
			SPE & Material & T & $\lambda$(nm) & $\gamma_\text{QD}$(meV) & $m_\text{eff}(m_e)$ & $r_0$(nm) & $E_b$(meV) & $\tau_l$(ns) \\ \hline
			\multirow{3}{*}{\tabincell{c}{Collidal\\ perovskite\\ QD}} & $\text{CsPbCl}_3$ & \multirow{3}{*}{\tabincell{c}{LT$\sim$RT}} & \tabincell{c}{400$\sim$450} & \multirow{3}{*}{\tabincell{c}{$\sim$1 (6K);\\ $\sim$100 (RT)}} & $0.092^\star$ & $5^\star$ & $75^\star$ & \\ \cline{2-2} \cline{4-4} \cline{6-9}
			& $\text{CsPbBr}_3$ & & \tabincell{c}{450$\sim$600} & & $0.073^\star$ & $7^\star$ & $40^\star$ & \tabincell{c}{$\sim$0.2 (6K);\\ $\sim$15.1 (RT)} \\ \cline{2-2} \cline{4-4} \cline{6-9}
			& $\text{CsPbI}_3$ & & \tabincell{c}{600$\sim$700} & & $0.06^\star$ & $12^\star$ & $20^\star$ & \\ \hline
			\multirow{4}{*}{2D material} & WSe2 & LT & $\sim$730 & $\sim$0.13 & & & & $\sim$1.79 \\ \cline{2-9} 
			& \tabincell{c}{WSe2\\ on pillar} & LT & \tabincell{c}{740$\sim$820} & $\sim$0.18 & & & & $\sim$2.8 \\ \cline{2-9} 
			& hBN & \tabincell{c}{LT$\sim$RT} & \tabincell{c}{$\sim$623\\ (77K)} & \tabincell{c}{$\sim$1.2\\ (77K)} & & & & $\sim$3.09 (RT) \\ \cline{2-9} 
			& \tabincell{c}{Moir\'{e}\\ exciton} & LT & $\sim$930 & $\sim$0.1 & & & ~420 & $\sim$12.1 \\ \hline
			\multirow{2}{*}{\tabincell{c}{Traditional\\ semiconductor\\ QD}} & GaAs & LT & \tabincell{c}{700$\sim$800} & $\sim$0.15 & $0.056^\star$ & $\sim$20 & $75^\star$ & $\sim$0.45 \\ \cline{2-9} 
			& CdSe & \tabincell{c}{LT$\sim$RT} & \tabincell{c}{470$\sim$620} & \tabincell{c}{$\sim$0.12 (10K);\\ $\sim$87 (RT)} & $0.12^\star$ & $\sim$11.4 & $110^\star$ & \tabincell{c}{$\sim$500 (10K);\\ $\sim$32 (RT)} \\ \hline
		\end{tabular}
		\caption{Parameters of various single photon emitters. Here SPE represents single photon emitter, T is the working temperature, $\lambda$ is the wavelength of single photon emission, $\gamma_\text{QD}$is the linewidth of single photon emission, $m_\text{eff}$ is the effective mass of exciton, $m_e$ is the mass of electron, $r_0$ is the exciton Bohr diameter, $E_b$ is the exciton binding energy, and $\tau_l$ is the lifetime of single photon emission. LT represents low temperature and RT represents room temperature. The data with $\star$ are the results of theoretical calculation.}
	\end{table}

	WSe$_2$ SPEs induced by strain in the monolayer have been reported with promising spectral properties. The strain can be generated with randomly distributed substrate roughness and more importantly can be engineered at will using nanostructured pillars. In this approach, the deterministic transfer of the monolayers onto nanostructured pillars has been reported with promising single photon emitter yield. While single pillar sites on a sub-100 nm scale are accessible utilizing negative resist with standard e-beam lithography procedure, challenges appear when aligning the pillar sites into closely packed (distance smaller than 20 nm) SSH chains, where proximity effect should be mitigated through Monte-Carlo simulation on dose distribution. Targeted transfer of monolayer WSe$_2$ onto the pillars with post-processing steps like thermal annealing to enhance adhesion can promote arrayed single photon emission from WSe$_2$.
	
	For the perspective on the placement of colloidal perovskite nanocrystals into closely spaced lattice sites, efforts can be devoted to the precise filling of solution-processed colloidal quantum dots into substrate pre-patterned into SSH hole chains through lithography and dry etching technique. In this scenario, external force-assisted occupancy of the patterned sites has been investigated, for example, through capillary force, electrostatic interactions, optical forces, and scanning probes technique, wettability contrast between different areas of the substrate. Furthermore, appropriate dilution of the colloidal nanocrystal solution before spin-casting can make it possible for the single sites occupied by a single quantum dot.
	
	GaAs/InAs quantum dots are typically grown in the Stranski-Krastanov mode via molecular beam epitaxy or metal-organic chemical vapor deposition. However, the strain in the heterostructures induces the formation of islands and subsequent nucleation and assembly of quantum dots distributed randomly on the substrate. To overcome the random position issue, site-controlled growth has been investigated through pre-patterning of the semiconductor substrate before epitaxy, thus guiding the nucleation of quantum dots on the patterned sites during growth. Such site-directed growth is promising for aligning GaAs/InAs quantum dots into desired lattice chains.
	
	To experimentally probe the linewidth narrowing effect, coupling with microcavities is encouraged in order to enhance spontaneous emission rate through Purcell enhancement, minimize dephasing, and increase extraction efficiency. To excite the edge state in the bandgap of around 5 meV, pumping on the entire SSH chain with resonant excitation is preferred, avoiding exciting bulk states and also manipulating the fundamental exciton state coherently. In contrast, in the non-resonant excitation scheme, the scattering time to ground exciton state limits the indistinguishability. Challenges in the resonant excitation scheme are separating the pump and fluorescence signal, which can be achieved with a cross-polarization method, filtering out excitation laser signal in the collection path. In addition, topological nontriviality is manifested in the emission on the edge site, so emission should only be collected on this site. In this respect, harnessing the merits of nanophotonics to manipulate wave propagation and modify local electromagnetic field and radiation pattern on the nanoscale, we can interface the waveguiding structures between the detector and the edge site, making the collection solely on the edge site accessible.

\end{document}